# Mass uptake during oxidation of metallic alloys: literature data collection, analysis, and FAIR sharing

Saswat Mishra, Sharmila Karumuri, Vincent Mika, Collin Scott, Chadwick Choy, Kenneth H. Sandhage, Ilias Bilionis, Michael S. Titus, Alejandro Strachan

Affiliations

School of Materials Engineering, Purdue University, West Lafayette, Indiana 47907, USA

School of Mechanical Engineering, Purdue University, West Lafayette, Indiana 47907, USA

School of Materials Engineering and Network for Computational Nanotechnology, Purdue University, West Lafayette, Indiana 47907, USA

Corresponding author: strachan@purdue.edu

The area-normalized change of mass (Δm/A) with time during the oxidation of metallic alloys is commonly used to assess oxidation resistance. Analyses of such data can also aid in evaluating underlying oxidation mechanisms. We performed an exhaustive literature search and digitized normalized mass change vs. time data for 407 alloys. To maximize the impact of these and future mass uptake data, we developed and published an open, online, computational workflow that fits the data to various models of oxidation kinetics, uses Bayesian statistics for model selection, and makes the raw data and model parameters available via a queryable database. The tool, Refractory Oxidation Database (https://nanohub.org/tools/refoxdb/), uses nanoHUB's Sim2Ls to make the workflow and data (including metadata) findable, accessible, interoperable, and reusable (FAIR). We find that the models selected by the original authors do not match the most likely one according to the Bayesian information criterion (BIC) in 71% of the cases. Further, in 56% of the cases, the published model was not even in the top 3 models according to the BIC. These numbers were obtained assuming an experimental noise of 2.5% of the mass gain range, a smaller noise leads to more discrepancies. The RefOxDB tool is open access and researchers can add their own raw data (those to be included in future publications, as well as negative results) for analysis and to share their work with the community. Such consistent and systematic analysis of open, community-generated data can significantly accelerate the development of machine-learning models for oxidation behavior and assist in the understanding and improvement of oxidation resistance.

## Introduction

Machine learning holds great promise in materials science, with the potential to accelerate the optimization of existing materials and the discovery of new ones [1-10]. A major roadblock to this approach is the fact that most of the experimentally generated data is not findable, accessible, interoperable, and reusable (FAIR) nor ready for machine learning (ML). Thus, most of the demonstrations of ML-driven materials discovery involve relatively fast and inexpensive simulations [11-14] or experiments [15-17]. However, in many critical applications, experiments are expensive and time-consuming, and a need exists to provide legacy experimental results in electronic form for ML analysis. An important related need is the ability to easily integrate new data into a database with legacy data. Furthermore, the majority of the negative (but correct) results collected during research are not included in final publications and, thus, do not contribute to the community's knowledgebase. These needs are particularly pressing in the development of advanced materials for use in extreme conditions due to the complexity and expense of the experiments involved. For example, assessing the performance of metallic alloys for high temperature applications requires time-consuming, high-temperature mechanical and oxidation tests; the latter require, at minimum, one hour and can last for days to weeks [18]. In addition, oxidation testing data is challenging because specimen dimension and surface preparation methods are not standardized, units for surface area-normalized mass gain and oxidation rate coefficients are varied, and raw data is rarely available.

The study of area-normalized mass uptake kinetics during oxidation has been a widely used means of evaluating the oxidation resistance of alloys at a given temperature and environment. The functional form that describes the time dependence of such mass change provides information that, when combined with structural and chemical analyses of the oxidation product(s) and the metallic alloy, may be used to evaluate potential mechanisms controlling the oxidation process. For example, a linear change in mass per area with time (constant oxidation rate) could be consistent with a steady-state, diffusion-controlled oxidation process that occurs over a time-independent distance (e.g., mass transport through the oxide lattice of a dense inner zone of constant thickness within a thickening oxide scale containing outer porous/cracked zones [19], or mass transport through a gaseous boundary layer with a time-independent thickness [20]) or by a steady-state chemical reaction that occurs at a constant rate with time [21]. A decreasing rate of mass change per area with time that follows a parabolic rate law could result from an oxidation process controlled by mass transport through a thickening, dense, continuous, adherent external oxide scale [22] or by mass transport through a dense, continuous metallic phase within a thickening internal oxidation zone [23,24]. A rate of mass change per area with time that decreases faster with time than for parabolic kinetics (e.g., cubic or logarithmic rate laws) could result from an oxidation process controlled by mass transport along oxide grain boundaries in a thickening, dense, continuous, adherent, external oxide scale with oxide grain growth [25,26], or by a chemical reaction (e.g., chemisorption) that occurs at a decreasing rate with time [27]. Furthermore, because the primary mechanism controlling the rate of oxidation at a given temperature and in a given environment may change with time, a mass uptake curve over an extended time may follow two different rate laws.

These varied oxidation mechanisms further complicate efforts to collect and summarize oxidation data because multiple, non-standard methodologies are used to fit or characterize mass gain curves ranging from fitting by hand on linear-linear or log-log plots to categorizing curves by their appearance. Because of these approaches, mass gain coefficients and exponents are often not reported in literature sources. Despite these challenges, recent efforts have successfully collected significant data for Ni-, Fe-, and Ti- based alloys exhibiting parabolic oxidation kinetics [28,29] and more broadly high entropy alloys [30] and MAX phases [31]. However, specific oxidation rate laws have not been collected and analyzed using a standardized and consistent methodology.

In this paper, we collected oxidation kinetic data for metallic alloys from the literature, re-analyzed the raw data utilizing Bayesian statistics, and made the data, analysis tool, and results FAIR by building on the U.S. National Science Foundation nanoHUB's Sim2Ls infrastructure [32].

# 2. Data Collection and Analysis

## 2.1 Data collection

Oxidation mass gain curves of pure elements, binary alloys, ternary alloys, and high entropy alloys were collected from literature up to 2022. Google Scholar and Science Direct were searched using keywords including "oxidation", "high entropy alloys", and "refractory alloys" to identify potential data sources. To be included in the database, literature sources must have included the composition of the tested alloys and/or metals, total mass gain in units of mass per unit area, testing temperature, plots showing mass gain over time, or a table of interrupted oxidation experiments, surface preparation methods, and testing conditions (temperature and partial pressure of oxygen). We collected data if the alloy composition comprised approximately 50 at% or more refractory elements, and for mass gain experiments that had 4 or more data points. Special attention was given to collecting as much refractory high entropy alloy oxidation data as possible, but not all pure element, binary alloy, and ternary alloy were collected due to time constraints. In total, 330 continuous and 77 interrupted mass gain experiments were collected, representing 145 unique elements and alloys from 68 literature sources. Both interrupted and continuous mass gain curves were extracted using WebPlotDigitizer [33] software. For continuous mass gain data, curves were automatically extracted within WebPlotDigitizer using a spacing resulting in approximately 36 points per curve.

Nb, Ti, and Cr were the most common elements present in the database, as shown in Figure 1(a), and they were present in 140, 118, and 120 data curves, respectively. The distribution of metal alloy type (unary, binary, ternary, or high entropy alloy) is shown in Figure 2(b), where high entropy alloy represents alloys with four or more components.

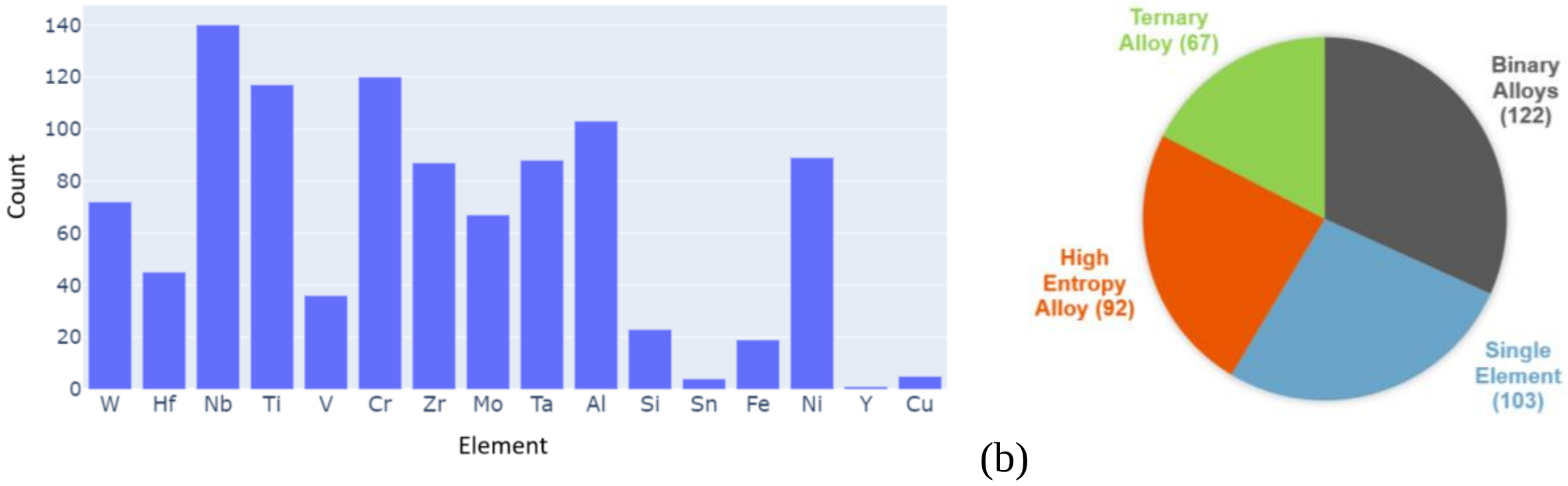

(a) (b)

*Figure 1. (a) Distribution of elements present within the database. (b) Distribution of alloy type within the database, with High Entropy Alloy representing alloys with four or more components.*

The temperature distribution of the data collected reveals that the most common oxidation experiment temperature was 1000 ºC, which represents 22% of the dataset, as shown in Figure 2(a). The majority of the experiments below 600 ºC and above 1300 ºC focused on pure elements, and all alloys were only tested between 600 ºC and 1400 ºC. Figure 2(b) shows the distribution of mass gain vs. temperature.

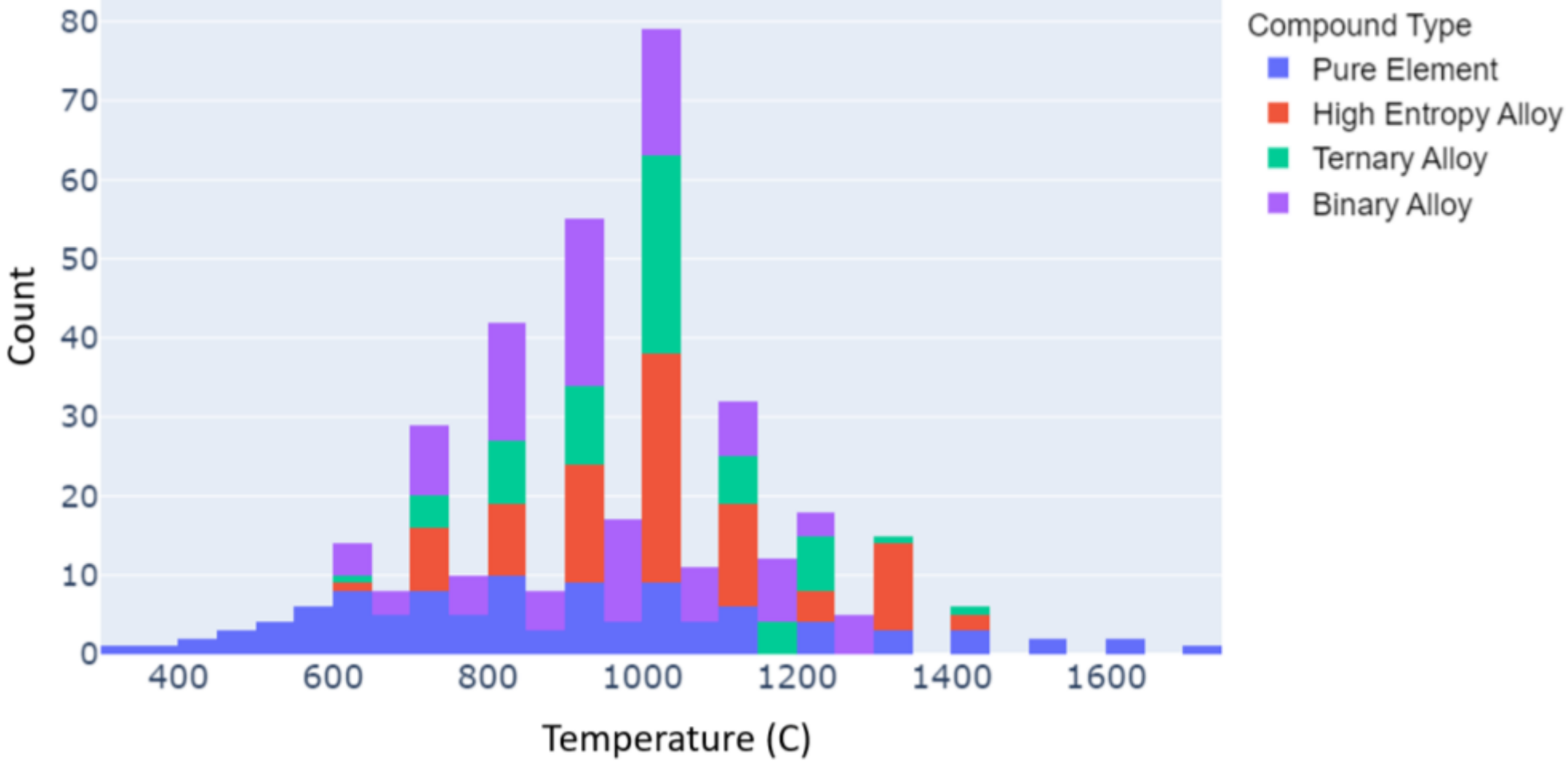

(a)

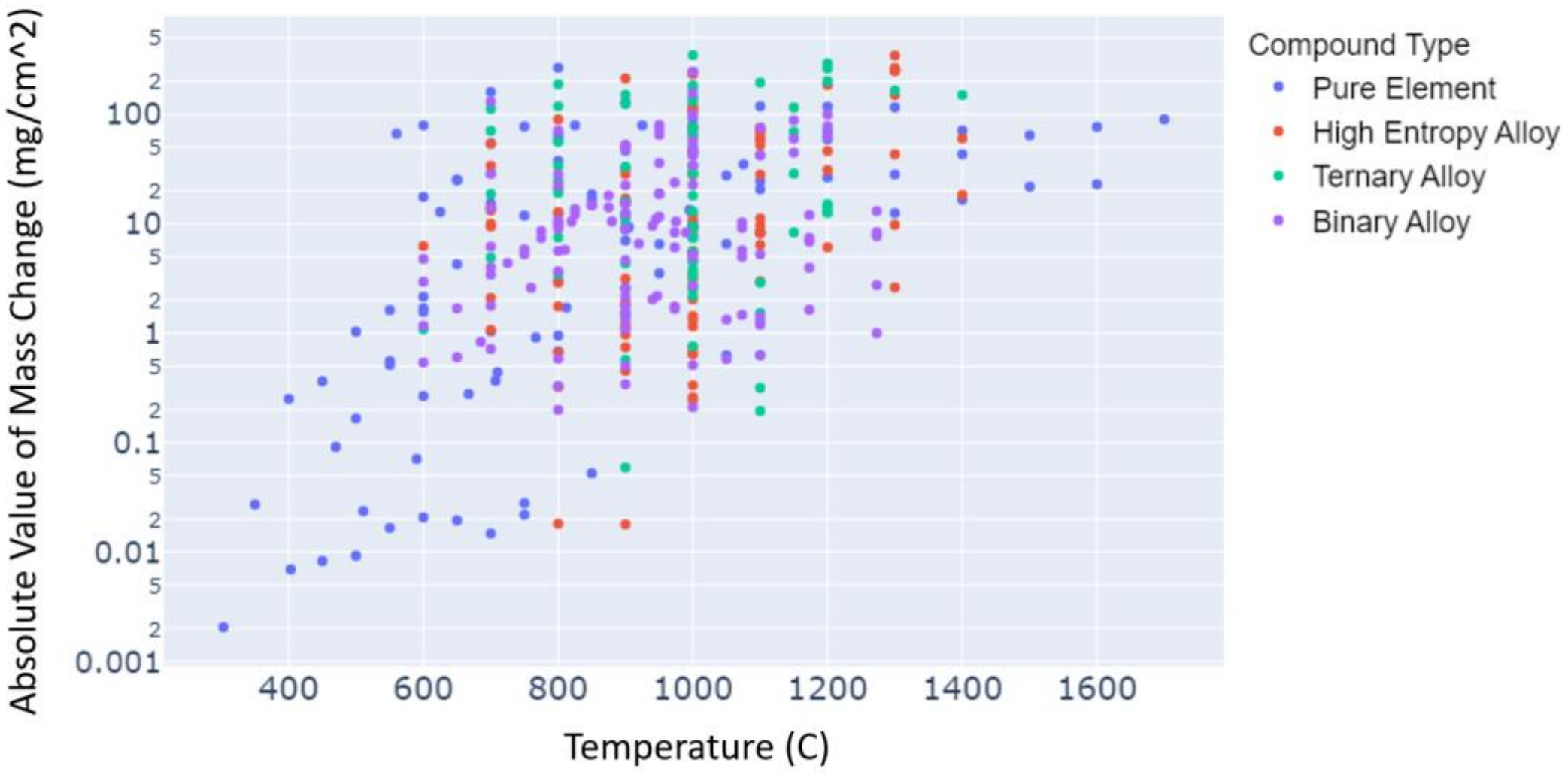

(b)

*Figure 2. (a)Temperature distribution of the mass gain data collected from literature (b) absolute value of mass change as a function of temperature color coded according to the number of elements in the alloy.*

## 2.2 Data analysis

Our goal is to develop a workflow that works without expert intervention or parameter tuning for most of the data published in the literature. To achieve this, we developed a Bayesian calibration method to parameterize and contrast various models to experimental oxidation mass gain vs. time curves. Depending on the underlying oxidation mechanisms, mass gain curves can follow a variety of functional forms. They can be written as:

$$\Delta m(t;\theta) = f(t;k,C)$$

where $\Delta m$ is the mass gain per unit surface area, $t$ is the oxidation time, and $\theta = (k, C)$ are the parameters of the model. The parameter $k$ is the mass gain rate constant and $C$ is an offset parameter. Table 1 lists six widely used single-mechanisms oxidation models considered here. As will be discussed next, both the mechanisms and functional form can change during oxidation due to microstructural and oxidation scale evolution.

| **Model type** | **Model id** $i$ | **Function** $f_i(t;k,C)$ |
|---|---|---|
| linear | 1 | $kt + C$ |
| quadratic | 2 | $kt^{1/2} + C$ |
| cubic | 3 | $kt^{1/3} + C$ |
| quartic | 4 | $kt^{1/4} + C$ |
| log | 5 | $k \log t + C$ |
| inv-log | 6 | $\frac{k}{\log t + C}$ |

Table 1: Single-behavior oxidation models considered in this work

In the collected experimental data, we observed that some alloys follow a particular mass gain behavior until a crossover time after which they switch to a different mass gain form. We constructed switching models to describe the situation by joining continuously two single-mechanism models, say $f_i(t; k_i, C_i)$ and $f_j(t; k_j, C_j)$ at time switch time point $t_c$. To transition from one model to the other continuously, we use the help of a sigmoid function scaled by a smoothness parameter $\gamma$ and centered at $t_c$, $\sigma(\gamma(t - t_c))$, see Fig. 3. The switch-behavior models have the form:

$$\Delta m_{ij}(t; \theta_{ij}) = f_i(t; k_i, C_i)[1 - \sigma(\gamma(t - t_c))] + f_j(t; k_j, C_j)\sigma(\gamma(t - t_c)).$$

The parameters of the model to be calibrated are $\theta_{ij} = (k_i, C_j, k_i, C_j, t_c)$.

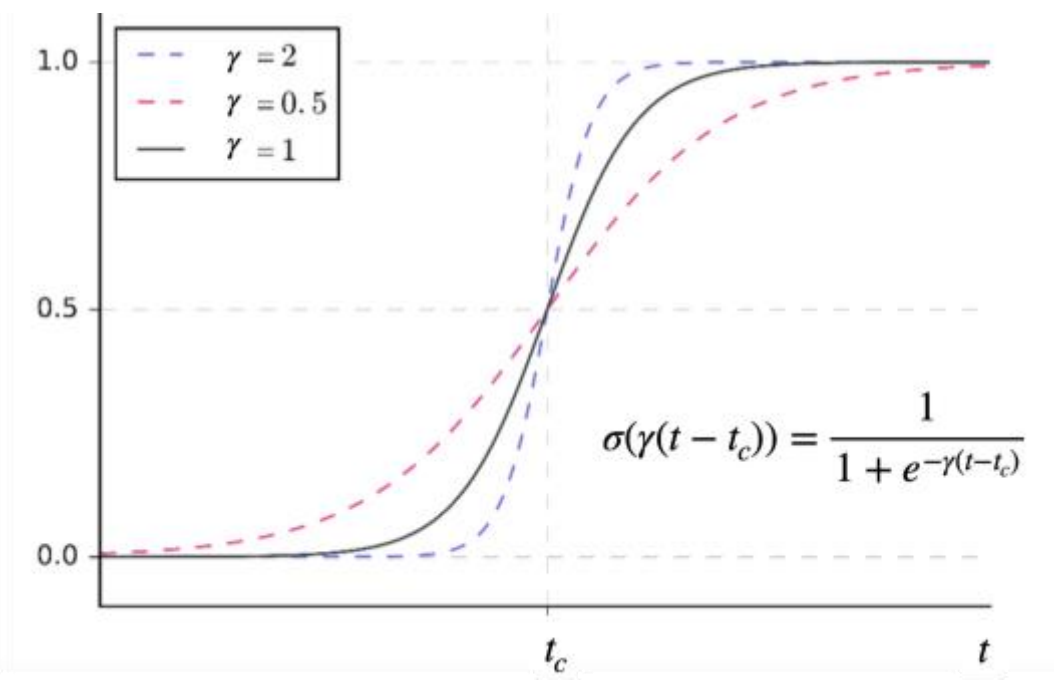

*Figure 3: Sigmoid function used to switch between regimes.*

As the sigmoid function varies between zero and one, this construction ensures that to the left of $t_c$ follows function $f_i(t; k_i, C_i)$, and to the right of $t_c$ it follows function $f_j(t; k_j, C_j)$. We fix the parameter $\gamma$ to a constant value of 100.

In total, each mass gain dataset is fitted against a total of 42 different models. These include the 6 single-behavior models of Table 1 and all 36 switch-behavior models (not that the switch can involve the same functional form with different parameters). Three-regime cases are not considered.

We formulate the parameter estimation problem in a Bayesian way:

a) Suppose we have experimental mass gain data $\mathcal{D} = (t_{\text{data}}, \Delta m_{\text{data}})$, where the times are $t_{\text{data}} = (t_{\text{data},1}, \dots, t_{\text{data},N})$, and the mass gains are $\Delta m_{\text{data}} = (\Delta m_{\text{data},1}, \dots, \Delta m_{\text{data},N})$. We assume independent measurements with Gaussian likelihood. For the switch-behavior models it is:

$$p\left(\Delta m_{\text{data}} \middle| t_{\text{data}}, \theta_{ij}, \lambda_{ij}\right) = \prod_{n=1}^{N} \mathcal{N}\left(\Delta m_{\text{data},\,n} \middle| \Delta m_{ij}\left(t_{\text{data},\,n}; \theta_{ij}\right), \lambda_{ij}^2\right),$$

where $\lambda^2$ is the noise variance of the experimental mass gain data, to be determined. A similar equation holds for the single-behavior models.

b) We assume that all model parameters are a priori independent. For the mass gain rate and the offset, we pick a zero-mean Gaussian prior with standard deviation of 10:

$$k_i \sim \mathcal{N}(0, 10^2), \qquad C_i \sim \mathcal{N}(0, 10^2).$$

For the noise standard deviation $\lambda$ we use a Half-Cauchy prior with hyper-parameter $\beta = 0.5$. In addition to this self-adjusting noise, our workflow enables the specification of a noise value as a fraction of the data range. When applicable, for the switch point $t_c$, we pick a uniform prior within 5%-95% of the oxidation time, i.e.,

$$t_c = a(\mathcal{D}) + \left(b(\mathcal{D}) - a(\mathcal{D})\right)z,$$

where $a = \min(t_{\text{data}}) + 0.05\,\Delta(t_{\text{data}})$, $\text{b} = \max(t_{\text{data}}) - 0.05\,\Delta(t_{\text{data}})$, and $\Delta(t_{\text{data}}) = \max(t_{\text{data}}) - \min(t_{\text{data}})$, and $z \sim \mathcal{U}(0,1)$ be the uniformly distributed parameter being actually estimated.

We find the best parameters by maximizing the posterior and we select the best overall model using the Bayesian information criterion (BIC). BIC scores the models based on their log-likelihood (ability to fit data) and complexity (number of parameters). The BIC score of a model j ($M_j$) is defined as [34]:

$$BIC_{M_j} = q\,log(N) - 2log(\hat{L}),$$

where, q = number of parameters in the model j, N = number of data points and $\hat{L}$ = likelihood of the data using model j.

Further using the BIC scores, we estimate the posterior probabilities for each of the models. We assume that prior probabilities $p(M_i)$ to be same for all the models. The likelihood of the data given a particular model can be computed through BIC scores [35] as,

$$p\left(\mathcal{D}\middle|M_j\right) = \exp\left(-\text{BIC}_{M_j}/2 + O(1)\right) \approx \exp\left(-\text{BIC}_{M_j}/2\right).$$

Using this relation, the required posterior probabilities of the models reduce to,

$$p\left(M_j\middle|\mathcal{D}\right) = \frac{p\left(\mathcal{D}\middle|M_j\right)p\left(M_j\right)}{p(\mathcal{D})}$$

$$= \frac{p(\mathcal{D}|M_j)\, p(M_j)}{\sum_{i=1}^{N_{models}} p(\mathcal{D}|M_i)\, p(M_i)}$$

$$= \frac{p(\mathcal{D}|M_j)}{\sum_{i=1}^{N_{models}} p(\mathcal{D}|M_i)}$$

$$= \frac{\exp\left(-\mathrm{BIC}_{M_j}/2\right)}{\sum_{i=1}^{N_{models}} \exp\left(-\mathrm{BIC}_{M_i}/2\right)}.$$

To exemplify our approach, consider the experimental mass gain data in Fig. 4 for a Hf1.0 alloy (circles) at 760 torr oxygen pressure at 1000$^o$C temperature [36] and the fits to the 42 models (lines). We note that many of the models describe the data quite accurately and one needs a rigorous approach for model selection. The resulting BIC scores are compared in Fig. 5 (a) and the associated model probabilities in Fig. 5(b). The preferred model (lowest BIC score) is shown in Fig. 6. Our approach indicates that the experimental data is best described with a cubic model until 1,243 minutes ($t_c$) followed by linear behavior, this model has probability of 0.8.

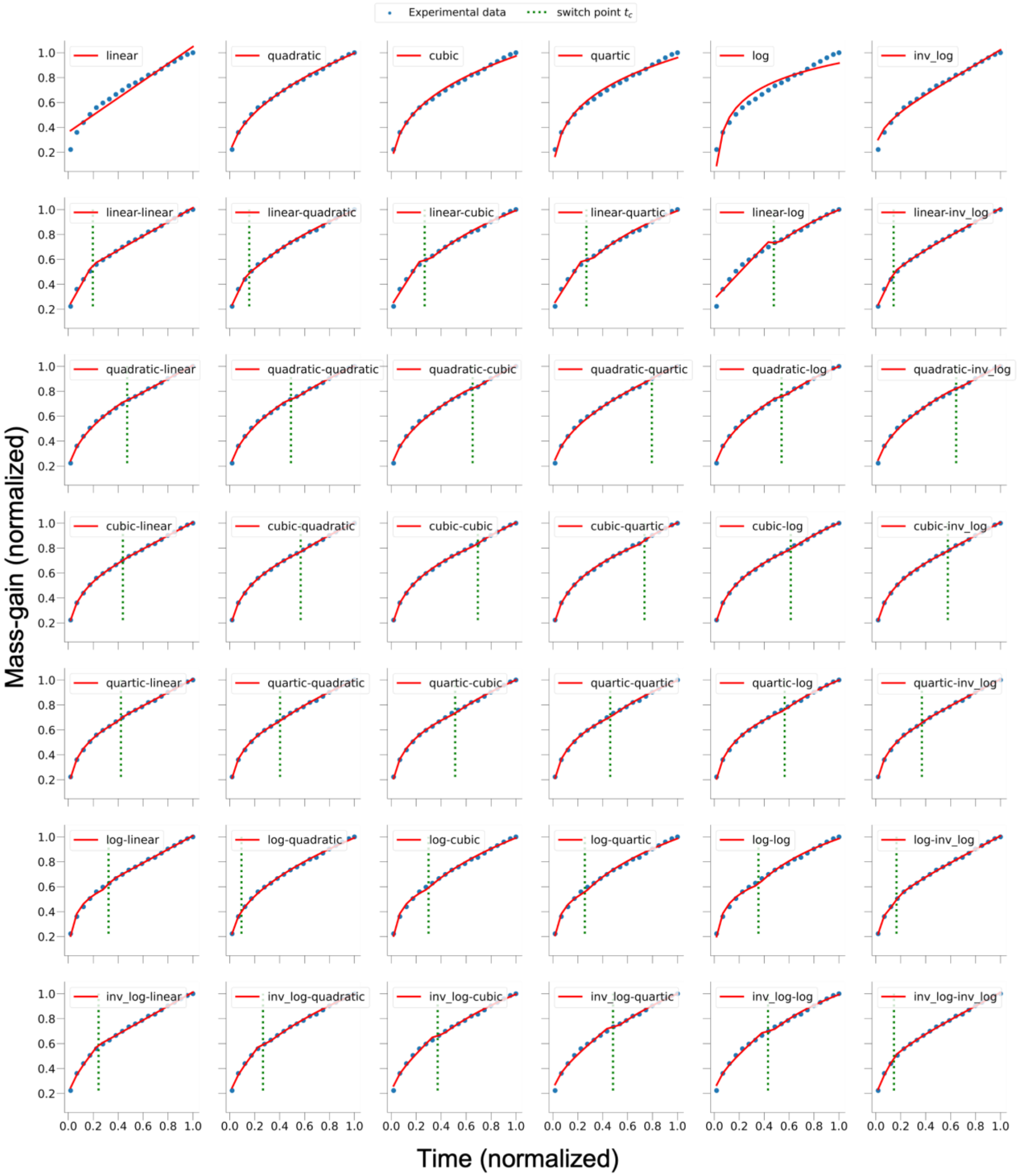

*Figure 4: Six single behavior models and 36 switch behavior models fitted to the data.*

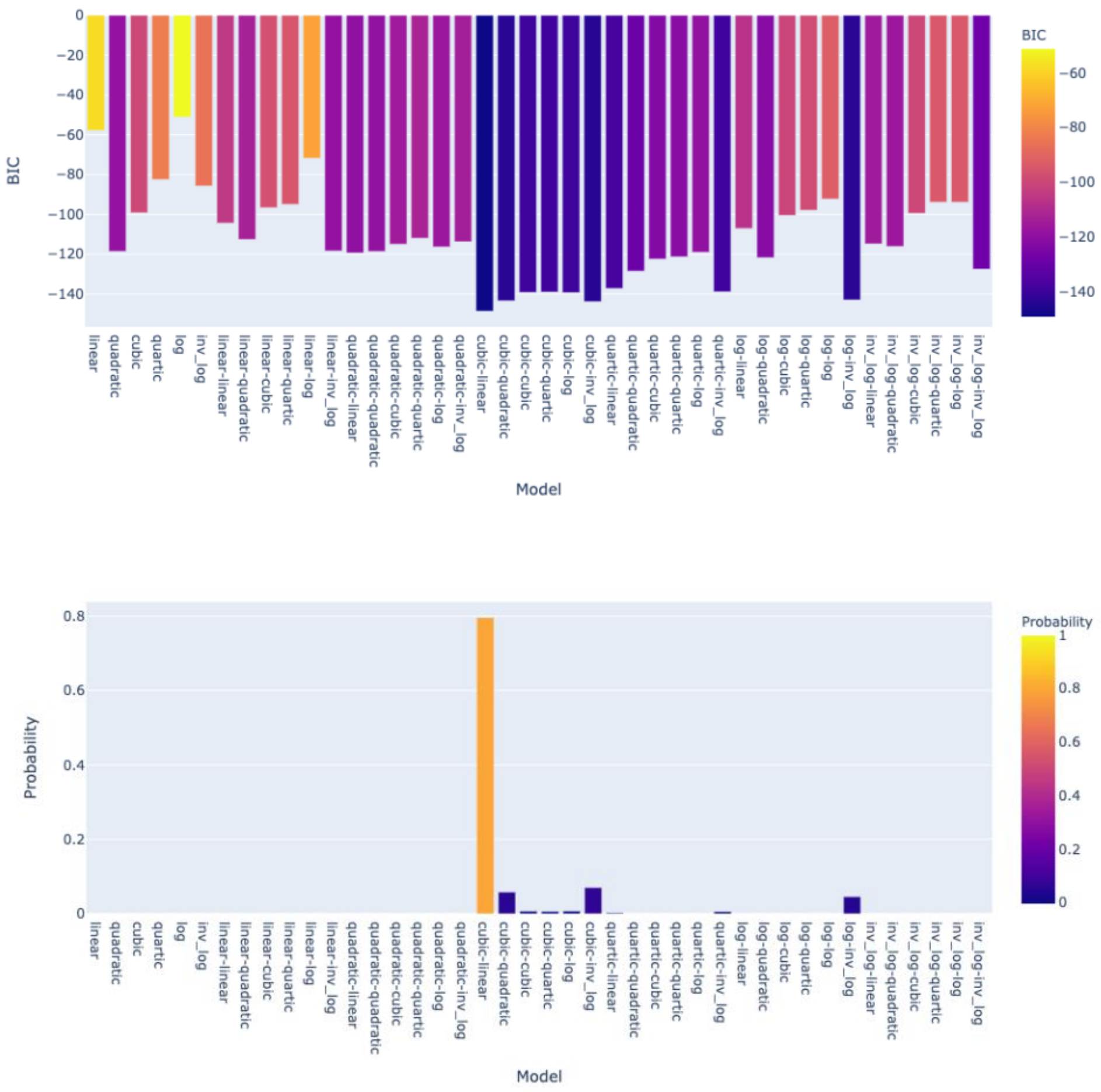

(a)

(b)

*Figure 5: BIC scores (a) and probabilities (b) of all the models for oxidation data of Hf1.0 alloy from [36]*

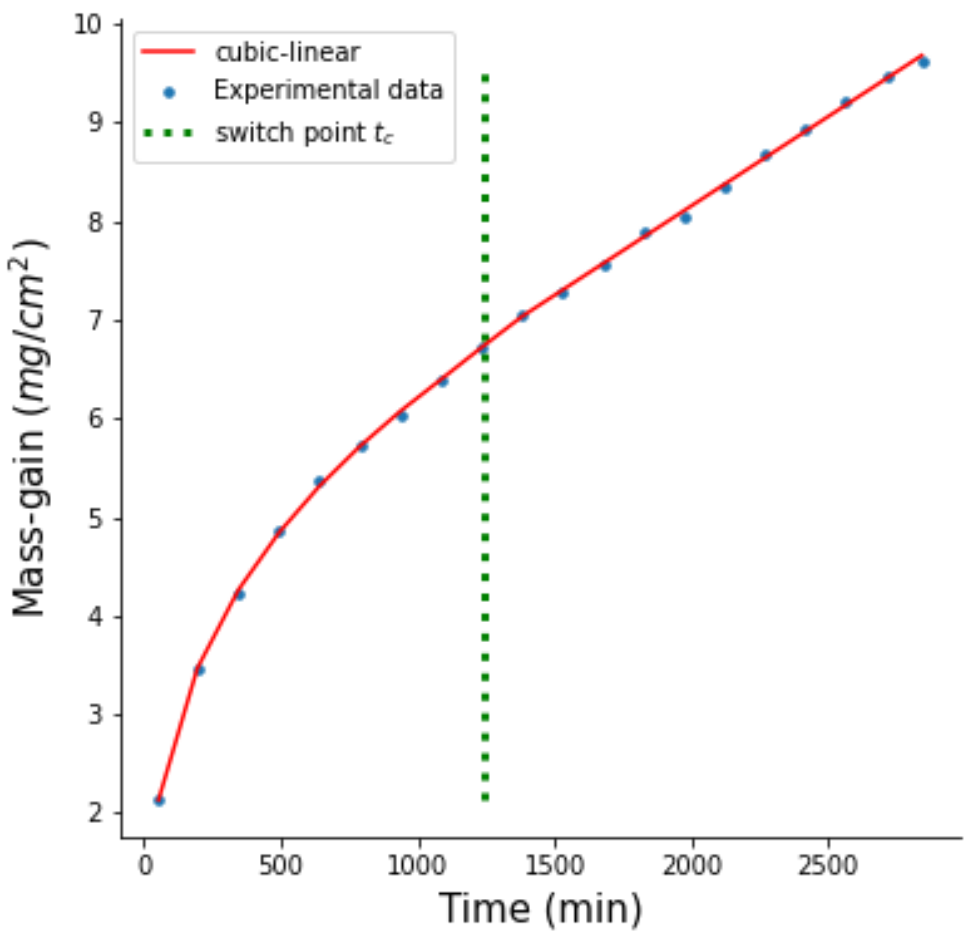

*Figure 6: Model with lowest BIC score.*

To conclude, our framework automatically fits oxidation mass gain curves and identifies the crossover points if any, and the corresponding oxidation behavior on either side of the crossover point by accounting for all the measurement uncertainties in the data.

## 3. Results

We collected raw data for oxidation mass gain vs. time for 407 alloys and the RefOxDB tool successfully calibrated and selected models for 405 of them. The authors reported models for 130 of these alloys (ignoring the alloys where the oxidation mass gain model was missing or was reported as a single mass gain curve exponent). As mentioned in the methods section, we collected data by extracting it from published plots which can add errors to the coordinates of the data points on top of the experimental errors. To account for this and the lack of reported experimental errors in some cases, we performed model calibration and selection using different noise levels: half-Cauchy, 1% of the mass gain range, and 2.5% of the mass gain range. As described in Section 2, the choice of error influences the likelihood term in BIC, with larger errors favoring simpler models.

Figure 7 shows the probability density function (PDF) and the cumulative distribution function (CDF) of the BIC ranking of the model selected in the original publication. The PDF shows the percentage of matches for each ranking while the CDF shows percentage of matches up to that rank. For a noise level of 2.5% of the mass gain range, we found that 29% of the models presented in the original paper matched the top BIC-selected model and 44% matched with one of our top 3 predictions. The number of matches reduces with the tightening of the noise assumed in the data for the analysis. We attribute this to researchers using their intuition for model selection and not performing an exhaustive exploration.

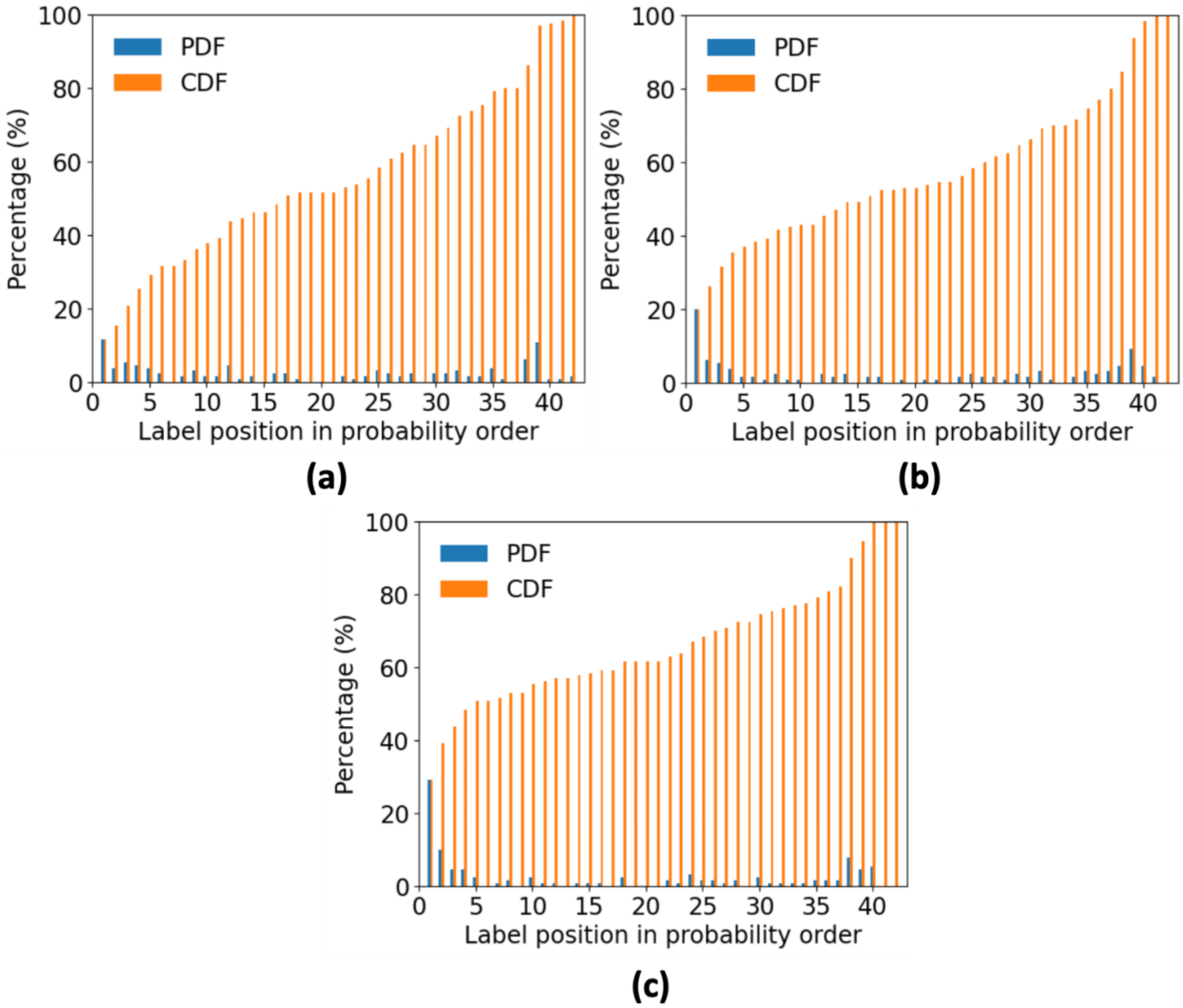

*Figure 7. position of the oxidation mass gain model prediction from literature in the probability sorted list of prediction from our model for (a) half-Cauchy noise (b) 1% noise (c) 2.5% noise. Blue bars show the PDF while the orange bars show the CDF.*

### 3.1 Examples of data analysis and comparison to literature results

Our analysis indicates cases where the BIC-predicted model agrees with the model selected by the authors and cases where the model reported in the literature was assigned a low probability. There were certain cases where even though the top model prediction did not match the literature label, the literature label was one of the top predictions from the model. We provide a few examples of all these cases here.

Figure 8 (a) shows a case where the literature reports a quadratic-linear model and our model also predicts the same. Figure 8 (b) shows the BIC score (lower is better) which clearly shows a low BIC score for the quadratic-linear case compared to the other models. Figure 8 (c) shows the probability of the models calculated from the BIC score.

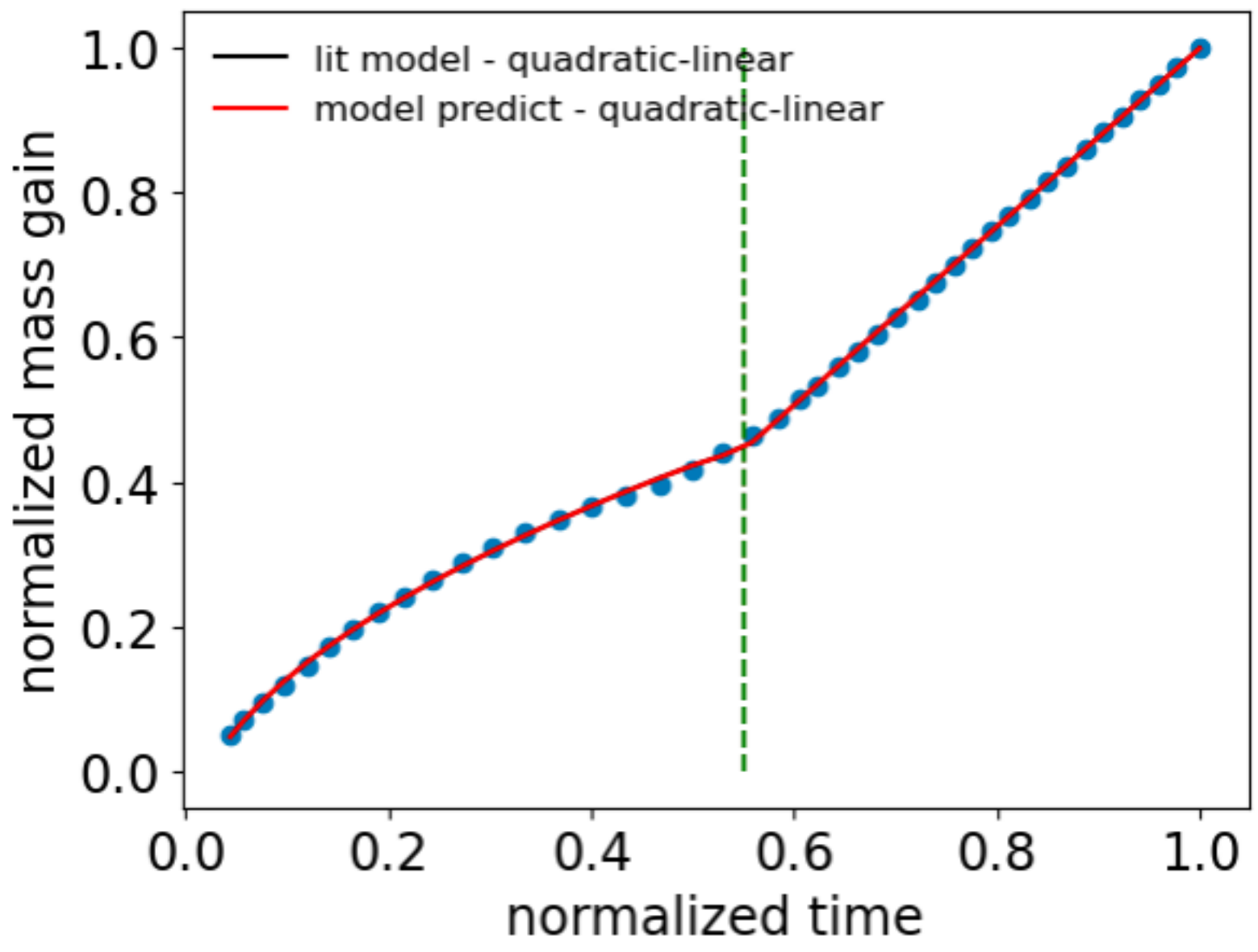

(a)

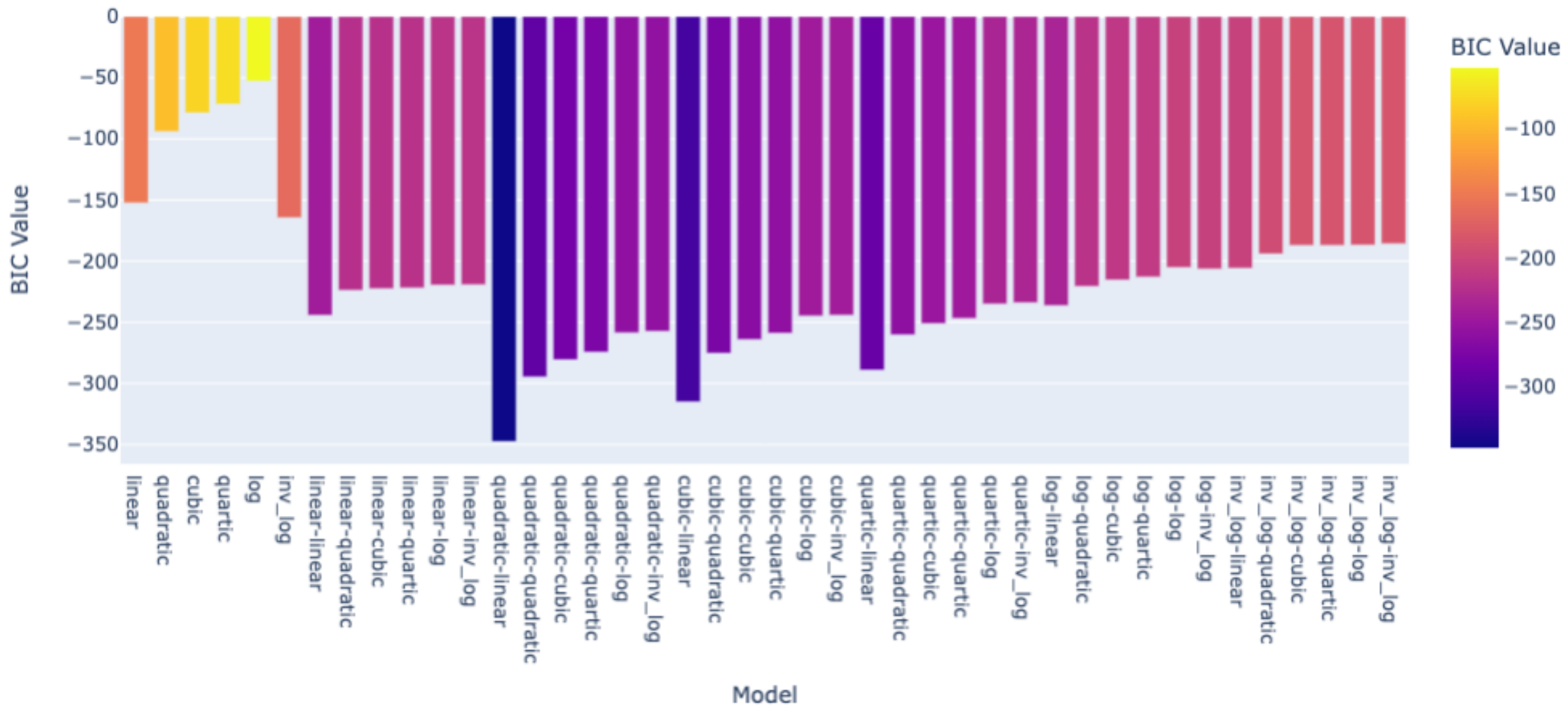

(b)

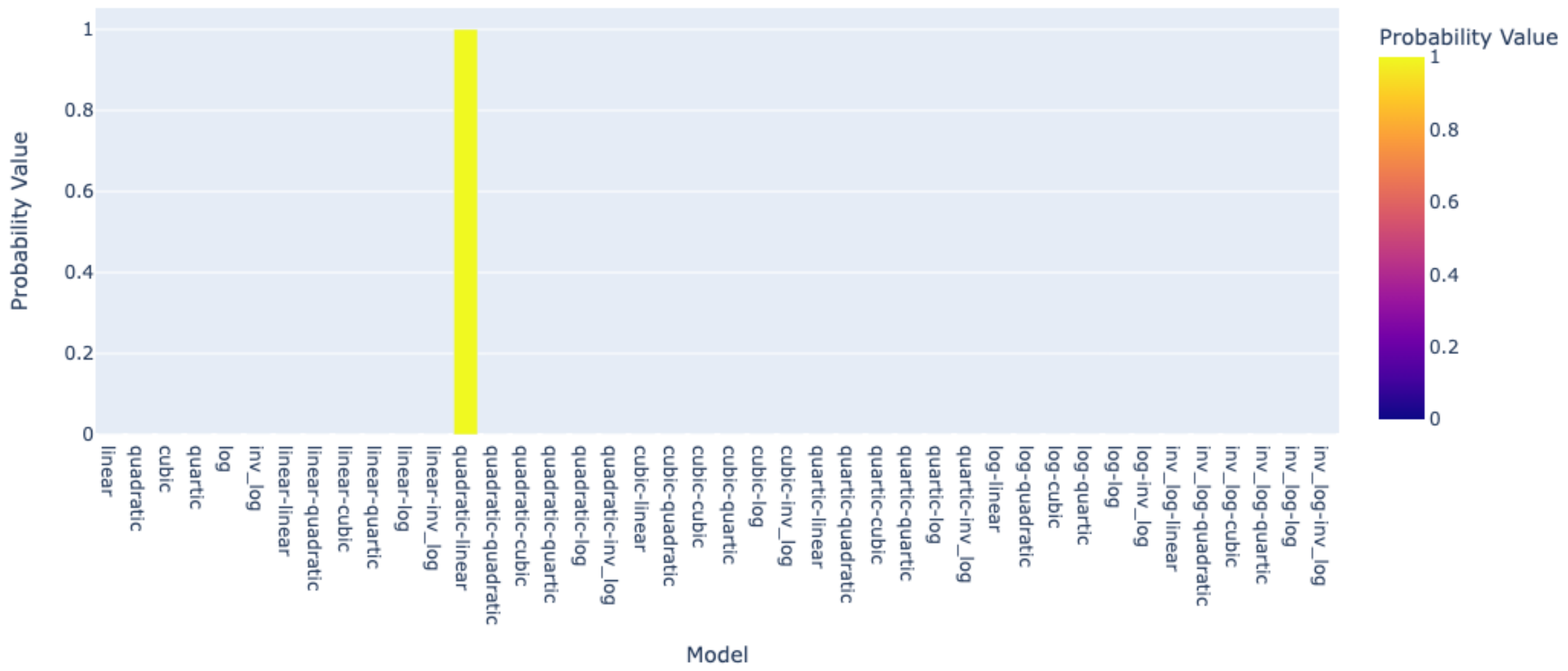

(c)

Figure 8. (a) mass gain vs time (b) BIC scores, and (c) probabilities in the case where the literature label and the model predictions match

Figure 9 (a) shows a case where the mass gain vs time curve was reported as quadratic-linear. Our model predicts it as cubic-linear. The curves look similar, but the model prediction fits better than the literature model. Figure 9 (b) shows the BIC scores, very close for quadratic-linear, cubic-linear, quartic-linear, and log-linear. As the probability is an exponential function of the BIC score, the difference shows up sharply in the probability plot in Figure 9 (c). Even though the mass gain vs time plot for quadratic-linear and cubic-linear look very similar, the model prediction assigns a significantly higher probability for the cubic-linear model over the quadratic-linear model. This example illustrates how a consistent framework gets rid of errors in prediction from eyeballing.

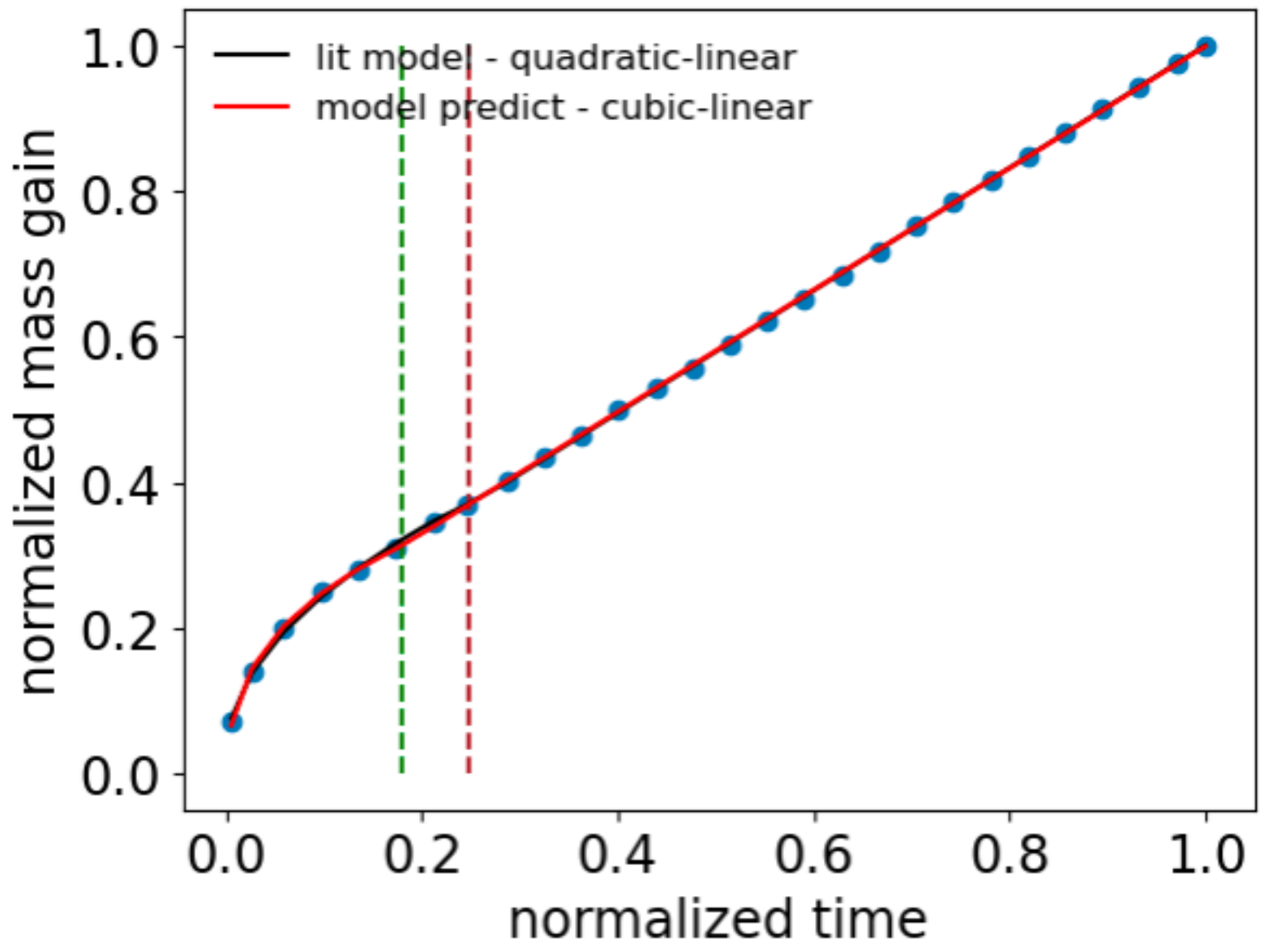

(a)

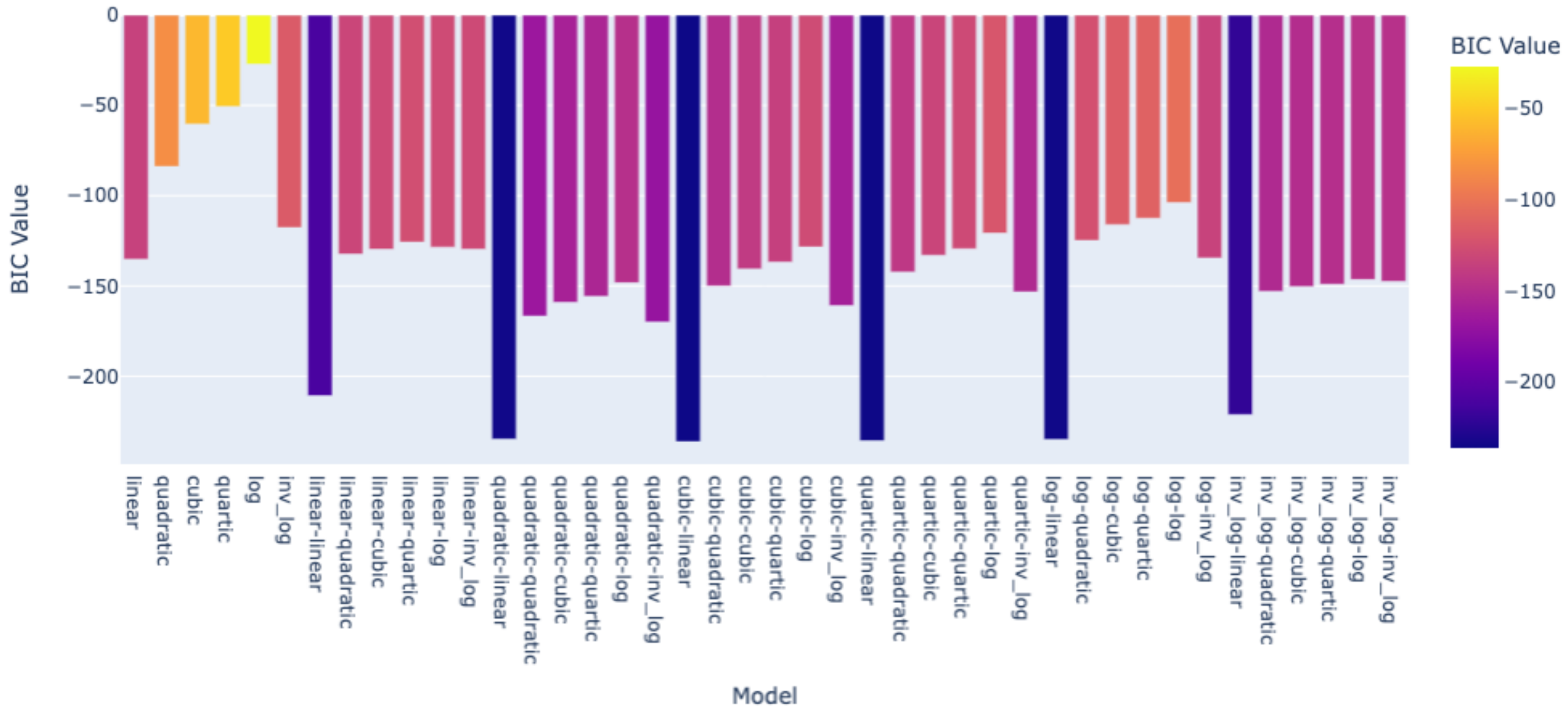

(b)

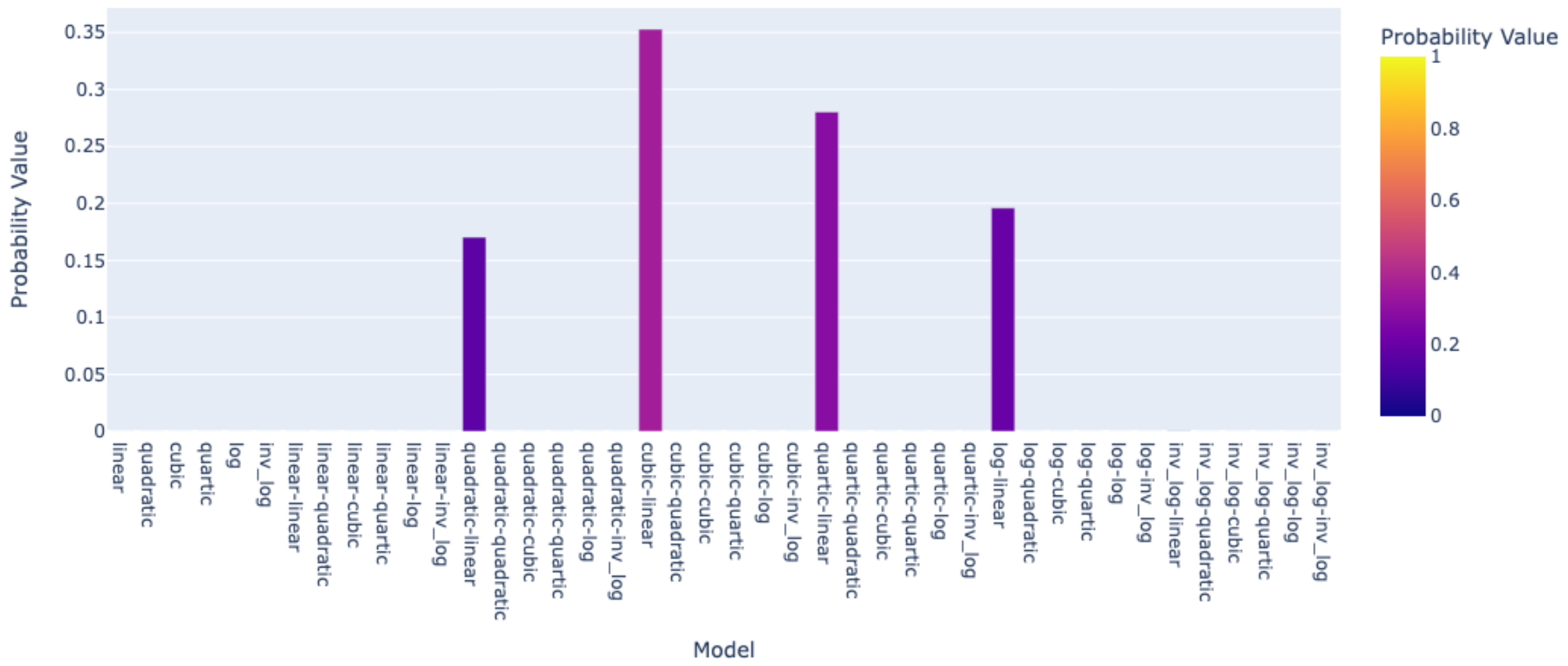

(c)

Figure 9. (a) mass gain vs time (b) BIC scores (c) probabilities in the case where the literature label and the model predictions do not match but the fits look similar

Figure 10(a) shows a case with a more distinct difference. Literature reports the mass gain curve as quadratic, whereas our model predicts it as quadratic-linear. The BIC score (Fig 10(b)) and the probability (Fig 10(c)) plot also show a better fit for the model prediction.

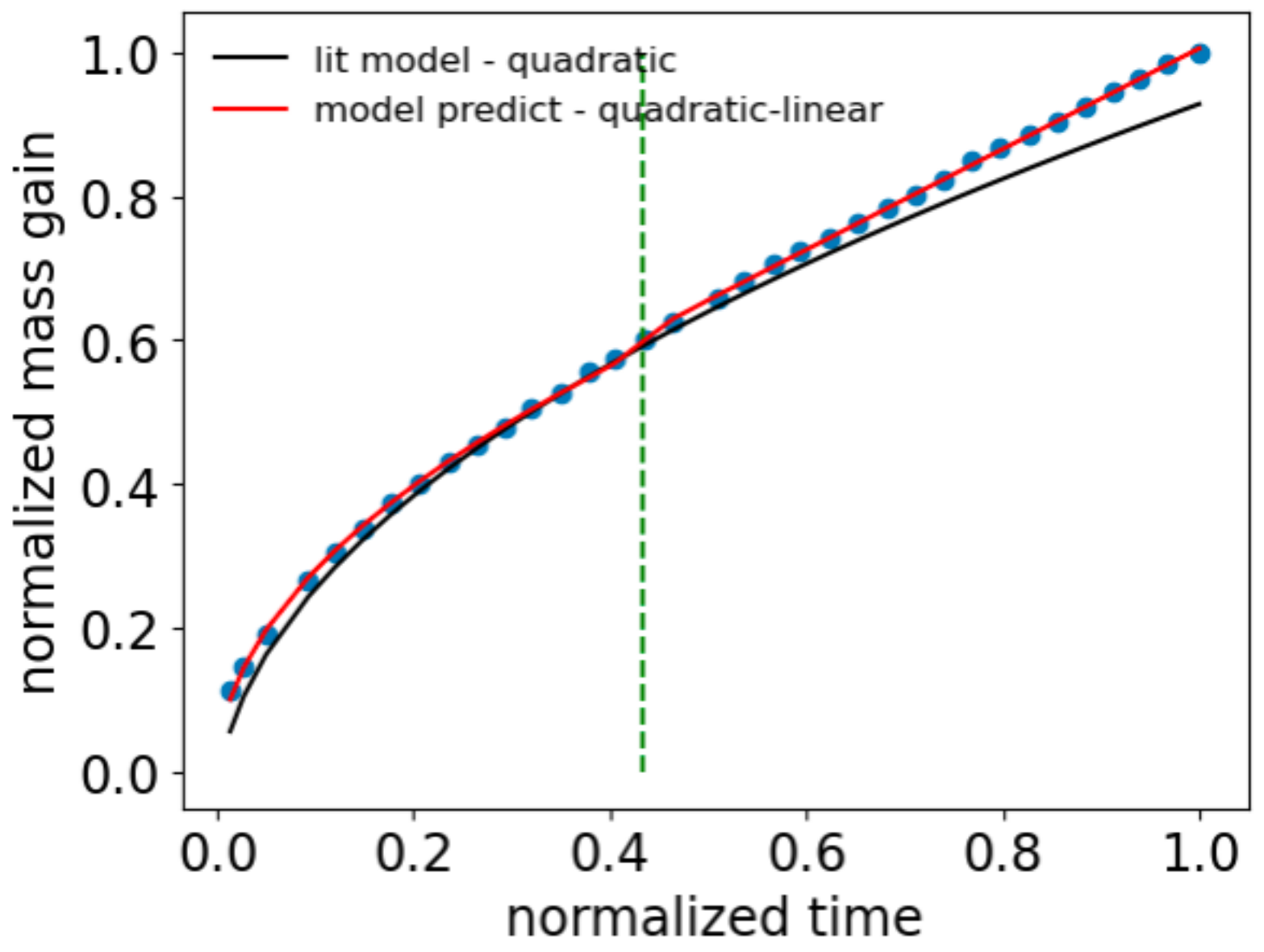

(a)

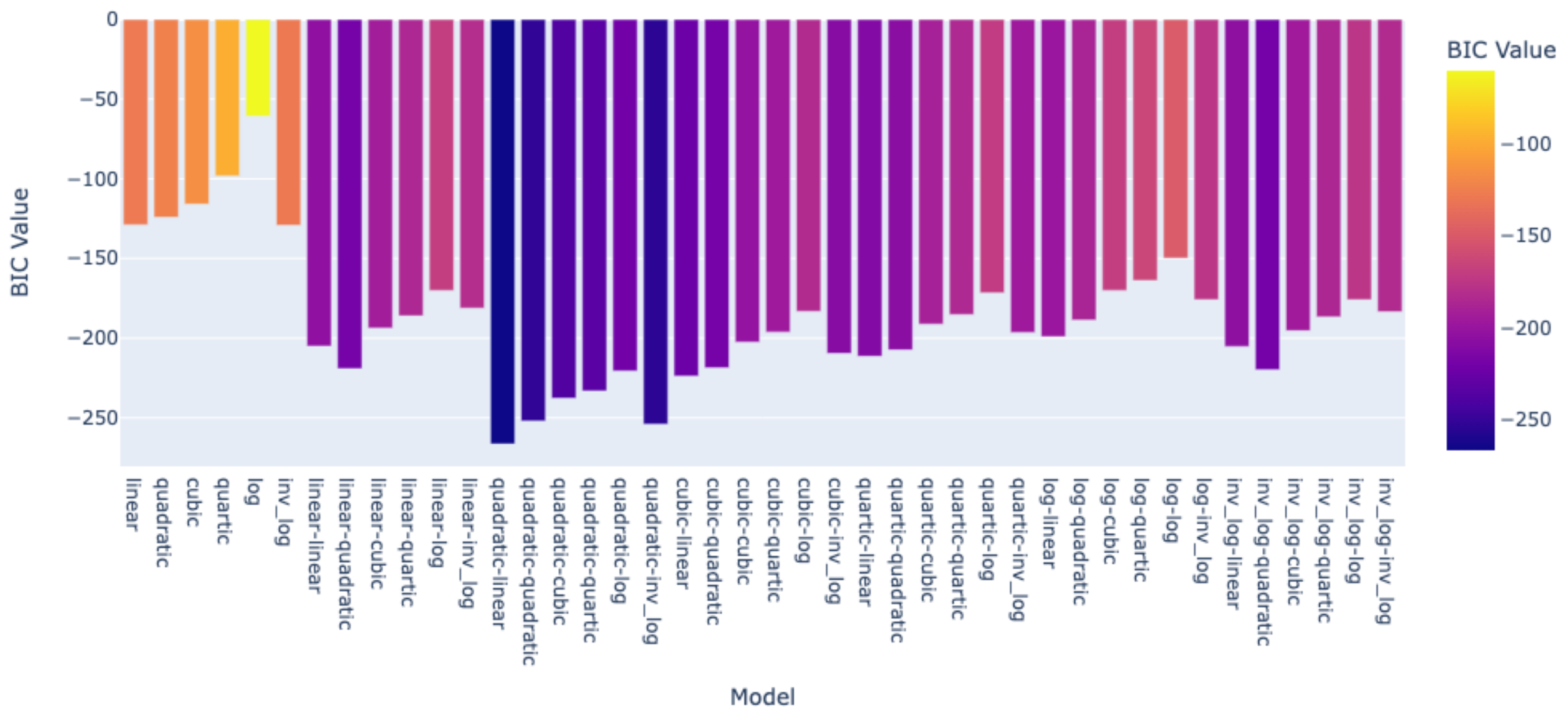

(b)

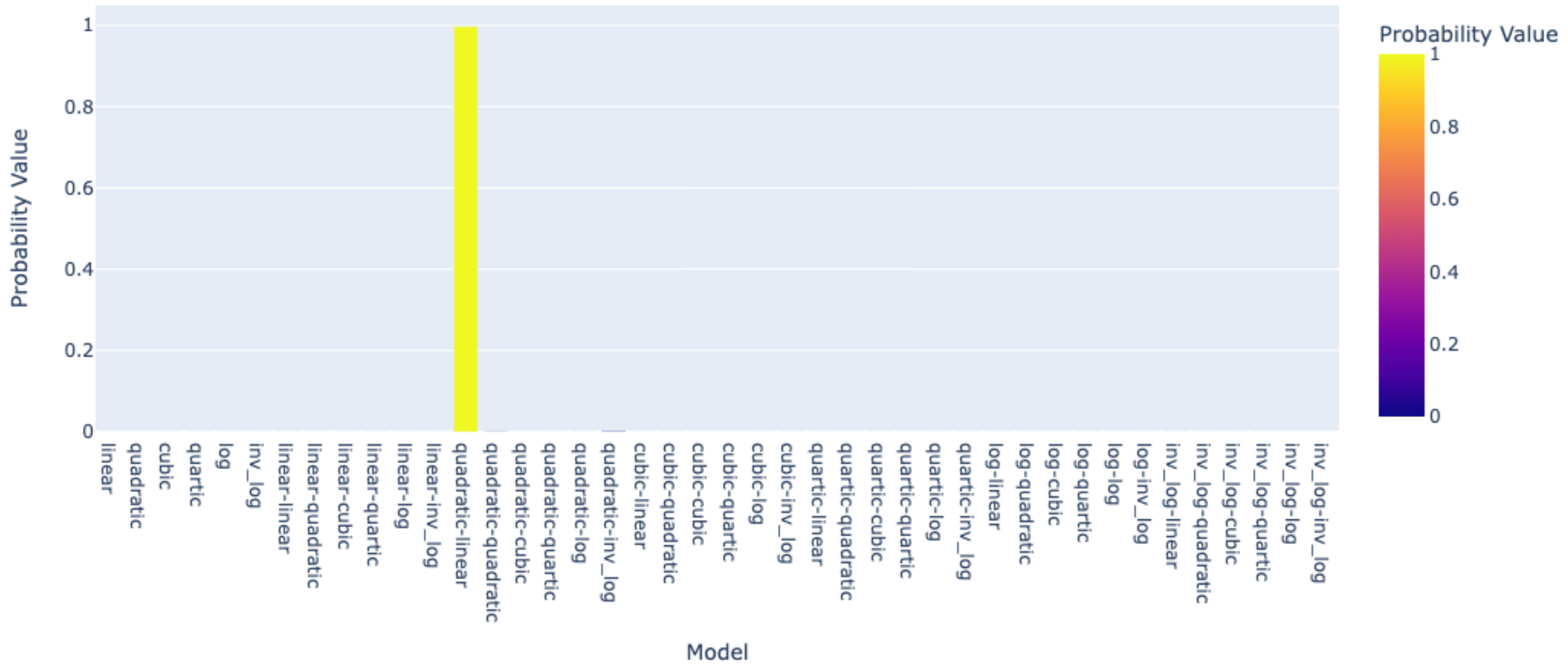

(c)

Figure 10. (a) mass gain vs time (b) BIC scores (c) probabilities in the case where the literature label and the model predictions do not match well, and the fits do not look similar

To examine how the noise influences the prediction capacity of the model, we compare the results of different noise parameters on the model prediction in Figure 11. It replicates Figure 9 and the best fits for both the noise cases (half Cauchy and 2.5%) are the same (shown once in Figure 11 (a)). However, when we change the noise parameter from the default half Cauchy distribution to an error of 2.5%, the probability of the other models (lower in Fig 9(c)) increases and become comparable to that of the predicted cubic-linear model. This is expected as the model considers the error inherent in each data point, thus bringing the alternative models in closer alignment with one another.

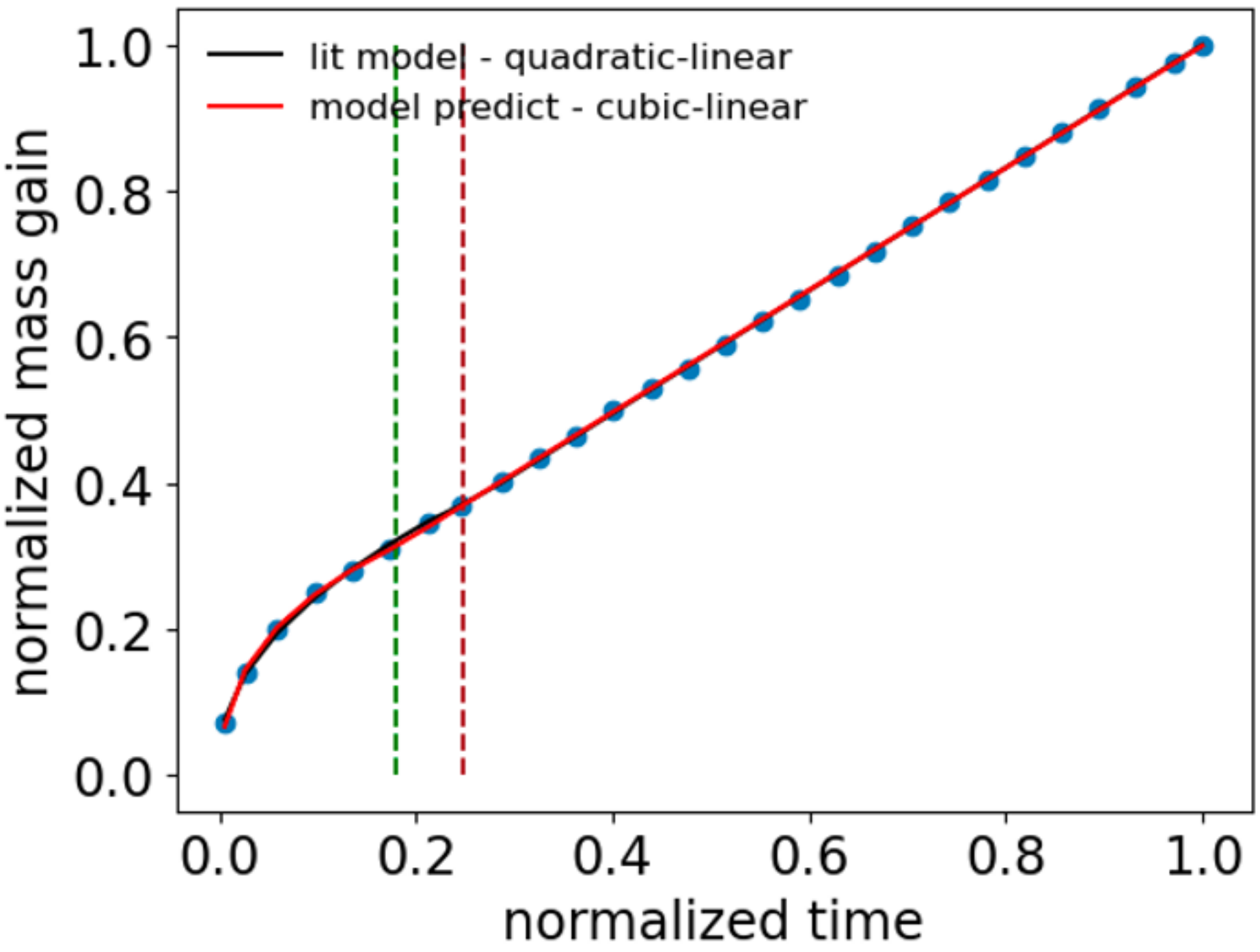

(a)

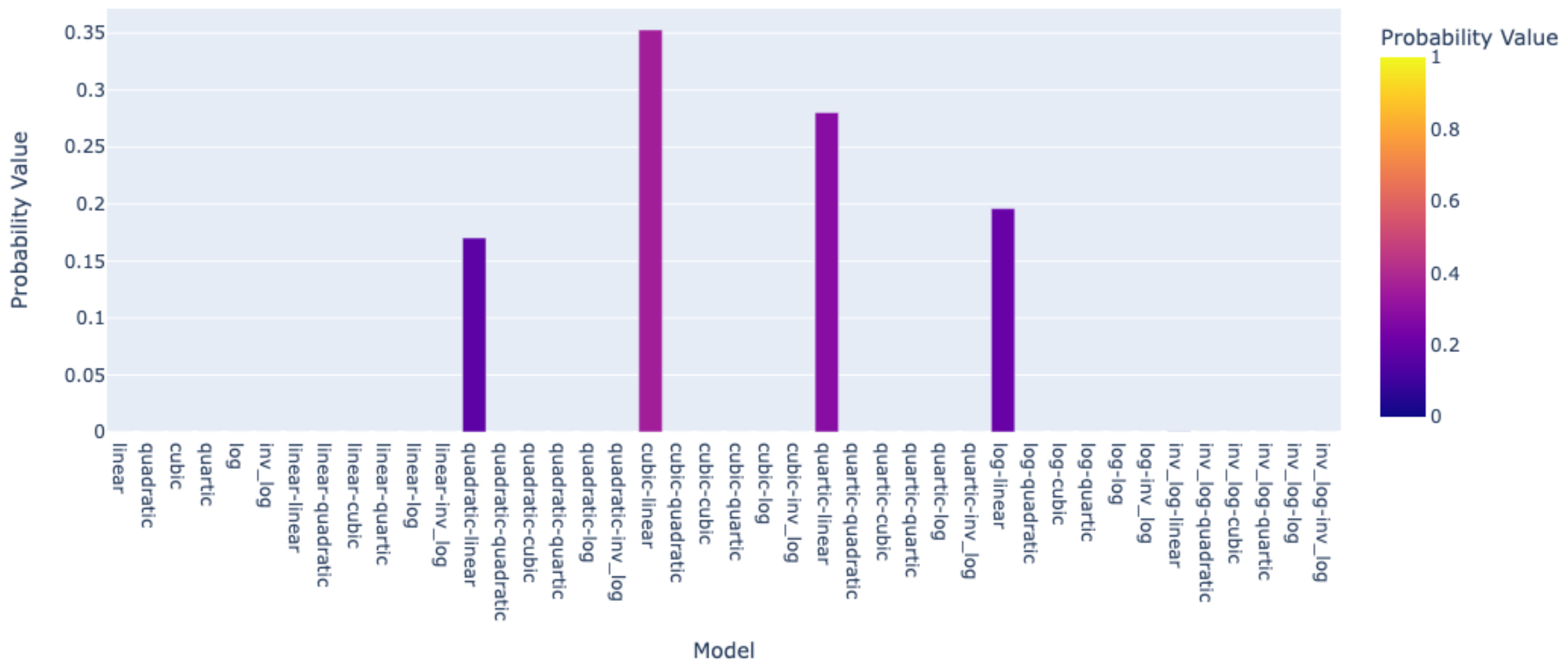

(b)

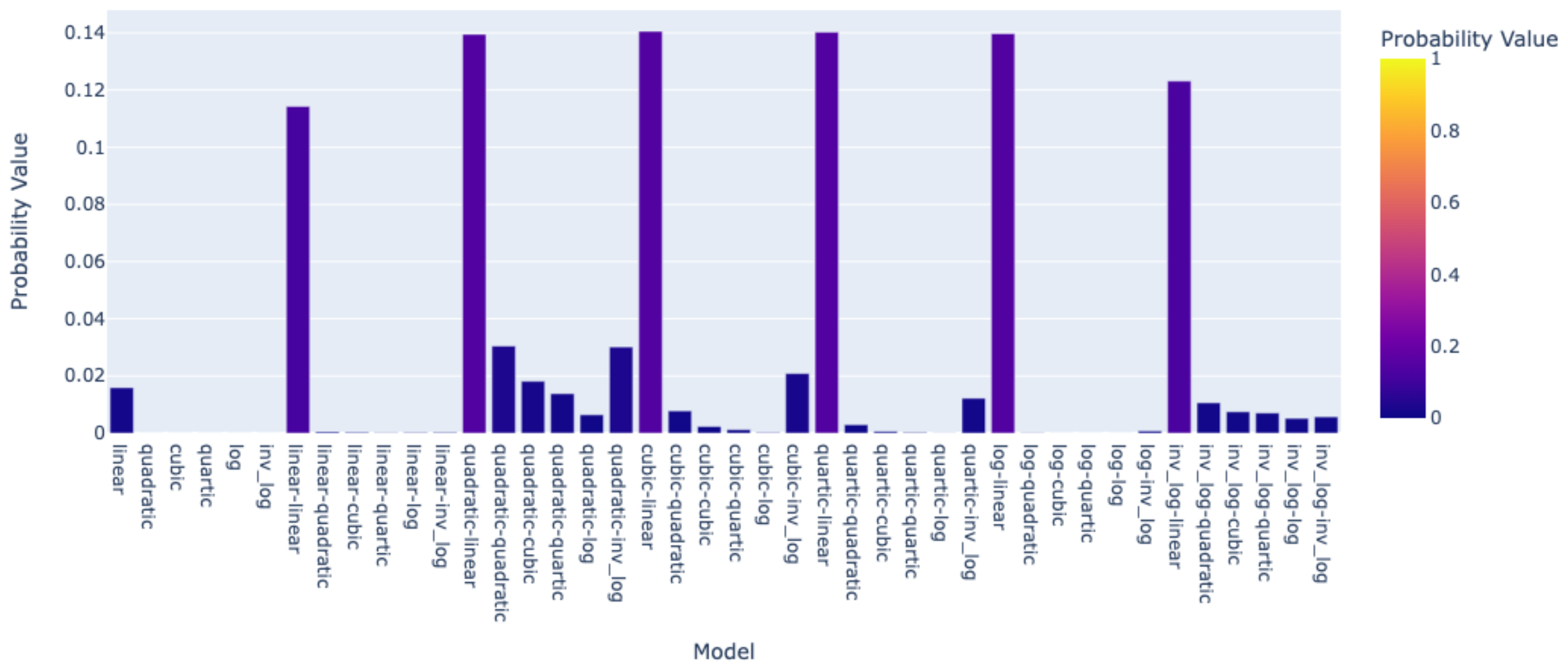

(c)

Figure 11. (a) mass gain vs time (b) probabilities for default noise, (c) probabilities for 2.5% noise

## 3.2 FAIR oxidation analysis tool and database

New data can be uploaded to the database by accessing the RefOxDB tool in nanoHUB (https://nanohub.org/tools/refoxdb/). The tool has a graphical user interface (GUI) option that lets the user input the fields manually while the batch upload option is better suited for cases where the user wants to upload and analyze multiple datasets. Table 2 shows the list of inputs and the outputs from the tool.

| **Inputs** | **Outputs** |
|---|---|
| Names | time |
| Affiliations | Mass gain vs time array |
| Reference | BIC plot |
| Year Publish | Probability plot |
| Reference ID | MAP dictionary |
| Chemical Formula | BIC dictionary |
| Formula ID | MSE dictionary |
| Temperature | $R_2$ dictionary |
| $p_{O_2}$ | Minimum BIC model |
| $p_{O_2}$ unit | Probability dictionary |
| $p_{other_gases}$ | Time switch point |
| Experiment method | Rate constant 1 |
| Surface preparation | Rate constant 2 |
| Heat treatment | |
| Fabrication method | |
| Gas flow rate | |
| Disintegration time | |
| Recession rate | |
| Oxide thickness | |
| Oxide time | |
| Oxides formed | |
| Phases present | |
| Grain size | |
| Shape | |
| Height | |
| Length | |
| Width | |
| Mass gain unit | |
| Time unit | |
| Mass gain vs time data | |

Table 2. Inputs and Outputs of the Refractory Oxidation Database tool. More details can be found in the supplementary material and the tool.

The database of can be accessed and the model predictions visualized in the same tool with the visualize option in the tool that lets the user query the database using filters.

## 4. Conclusions

In summary, we developed a computational workflow to calibrate oxidation kinetics models to experimental data and perform model selection using Bayesian statistics. The workflow takes experimental data, experimental conditions, and authorship information as inputs, performs calibration and model selection, and returns the optimized parameters for all 42 models and their BIC scores. The workflow is available for online simulation via the US National Science Foundation's nanoHUB which provides cloud computing services and storage. Implemented as a Sim2L [32], both the model and the data ingested and generated are FAIR. All data analyzed is available at Ref. 30 with a unique universal identifier (DOI) and researchers can add their experimental results accessing the online tool, Ref. 37

The analysis of 400+ datasets from the open literature indicates that often authors do not select the model that best fits their data and that experimental error is critical in model selection. Such consistent numerical analysis across a large number of experimental results and the raw data can help in developing effective machine learning models which in turn can help discover oxidation resistant alloys.

## Acknowledgments

We acknowledge the support from the U.S. National Science Foundation, DMREF program, under Contract No. 1922316-DMR, the Summer Undergraduate Research Program (SURF) at Purdue, and the computational resources from nanoHUB.

## Data Availability

The raw data required to reproduce these findings are available to download from https://nanohub.org/tools/refoxdb/. The processed data required to reproduce these findings are available to download from https://nanohub.org/tools/refoxdb/.